\documentclass[prb,twocolumn,showpacs,amsmath,amssymb,superscriptaddress]{revtex4-1}

\usepackage{graphicx}

\usepackage{dcolumn}
\usepackage{bm}
\usepackage{epsfig}
\usepackage{epstopdf}
\usepackage{color}

\begin{document}
\title{Transverse field muon-spin rotation measurement\\
of the topological anomaly in a thin film of MnSi}
\author{T. Lancaster}
\email{tom.lancaster@durham.ac.uk}
\affiliation{Centre for Materials Physics, Durham University, Durham,
  DH1 3LE, United Kingdom}
\author{F. Xiao}
\affiliation{Centre for Materials Physics, Durham University, Durham,
  DH1 3LE, United Kingdom}
\author{Z. Salman}
\affiliation{Laboratory for Muon Spin Spectroscopy, Paul Scherrer Institut, 5232 Villigen PSI, Switzerland}
\author{I.~O. Thomas}
\affiliation{Centre for Materials Physics, Durham University, Durham,
  DH1 3LE, United Kingdom}
\author{S.~J. Blundell}
\affiliation{Oxford University Department of Physics, Clarendon Laboratory, Parks
Road, Oxford, OX1~3PU, United Kingdom}
\author{F.~L. Pratt}
\affiliation{ISIS Facility, STFC Rutherford Appleton Laboratory,
  Chilton, Didcot, Oxfordshire, OX11~0QX, United Kingdom}
\author{S.~J. Clark}
\affiliation{Centre for Materials Physics, Durham University, Durham,
  DH1 3LE, United Kingdom}
\author{T. Prokscha}
\affiliation{Laboratory for Muon Spin Spectroscopy, Paul Scherrer Institut, 5232 Villigen PSI, Switzerland}
\author{A. Suter}
\affiliation{Laboratory for Muon Spin Spectroscopy, Paul Scherrer Institut, 5232 Villigen PSI, Switzerland}
\author{S.~L. Zhang}
\author{A.~A. Baker}
\author{T. Hesjedal}
\affiliation{Oxford University Department of Physics, Clarendon Laboratory, Parks
Road, Oxford, OX1~3PU, United Kingdom}

\begin{abstract}
We present the results of transverse-field muon-spin rotation 
measurements on an epitaxially grown 40~nm-thick film of MnSi on
Si(111) in the region of the field-temperature phase diagram where a
skyrmion phase has been observed in the bulk. We
identify changes in the quasistatic magnetic field distribution sampled by the
muon, along with evidence for magnetic transitions around $T\approx
40$~K and 30~K. Our results suggest that the cone phase is not the
only magnetic texture realized in film samples for out-of-plane
fields. 
\end{abstract}
\maketitle

There has been a flurry of recent interest in the physics of the skyrmion as an
example of a topological excitation in condensed matter.\cite{skyrm}
The simplest example of a skyrmion may be derived from a sphere
studded with radially directed arrows.  The skyrmion is formed via
the stereographic projection of the arrows onto a plane while keeping their orientations
fixed. 
The clearest evidence for the existence of the skyrmion is in the spin texture of magnetic systems and in
recent years a number of advances have demonstrated the existence, not only of magnetic
skyrmions, but also their ordering into a skyrmion lattice (SL).
\cite{Muhlbauer-2009,Munzer-2010,Yu-2010,Yi-2010a,Seki-2012a,Seki-2012b,Adams-2012,Seki-2012c,Langner-2014,omrani}
In bulk samples the SL has been observed in only a restricted region
of the applied field-temperature phase diagram of magnetic systems, known as the $A$-phase. However, it has been shown
that the SL phase is stabilized over an extended region of the phase
diagram in bulk samples that have been thinned.\cite{tonomura,Seki-2012a} This motivated the
search for skyrmions in epitaxially grown thin film systems. However,
the unambiguous identification of the SL in such samples has been
challenging.

When a magnetic field is applied along the
[111] direction  below the critical temperature $T^{\mathrm{bulk}}_{\mathrm{c}}=29.5$~K,  bulk MnSi
hosts four magnetically ordered phases, characterized by critical
fields\cite{Muhlbauer-2009} $B_{\mathrm{c}1}\approx 0.1$~T and $B_{\mathrm{c}2}\approx 0.5$~T.  Below a critical field $B_{\mathrm{c}1}$ the  spins order
helically with a $q$ vector  parallel   to   the  applied field;   for
$B_{\mathrm{c}1}<B<B_{\mathrm{c}2}$ the spins order conically; and when $B>B_{\mathrm{c}2}$ the spins
order ferromagnetically. In addition, for $B_{\mathrm{c}1}<B<B_{\mathrm{c}2}$ there
exists a wedge-shaped $A$-phase region close to $T^{\mathrm{bulk}}_{\mathrm{c}}$
(centered around $T=28$~K and an applied field $B_{\mathrm{app}}$=150~mT),
 which hosts the SL.\cite{Muhlbauer-2009, tonomura}
As with other B20 systems when compared to the bulk, the SL was reported to exist
over an extended region in thinned samples of thickness\cite{tonomura} $\approx
50$~nm
and also in nanowires.\cite{yu}
Subsequently, a report of topological Hall effect (THE)
measurements and Lorentz transmission electron microscopy (TEM) on
epitaxially grown thin-film samples suggested that the skyrmion phase was significantly
enlarged in field and temperature \cite{li} as might be presumed from
comparison between the phase diagrams of bulk samples and those that have been thinned. 
However, subsequent microscopy work\cite{Monchesky-2014}  challenged the notion that
skyrmions are present in the thin film samples, leading to the claim
that the extended phase region responsible for the THE response was,
in fact, the magnetic cone phase.\cite{meynell}
There have also been several experimental and theoretical investigations of thin-film MnSi
\cite{growth,karhu,wilson,wilson2,karhu2} suggesting, in particular, that the ground state 
magnetic structure propagates along the [111] direction.\cite{karhu}   Further, it was suggested that in
out-of-plane fields no first-order magnetic transition is observed
that would indicate the appearance of skyrmions and
that the cone magnetic structure is the thermodynamically stable phase for out-of-plane applied
magnetic fields with $B<B_{\mathrm{c2}}$.\cite{wilson2}

In view of the controversy, there is value in using alternative experimental
techniques to probe the physics of the magnetic field configurations
in thin film samples. 
To this end, we report the results of transverse-field muon-spin
rotation (TF $\mu^{+}$SR) measurements \cite{steve} on a thin-film sample of
MnSi. We probe the local magnetic field distribution in the region
giving rise to the reported anomalous topological response in Hall
measurements. Here we show that several discontinuous changes in the
field distribution are observed in the $T$=20--40~K
region suggesting that the magnetic structure changes significantly,
and indicating that the cone phase is unlikely to be the only magnetically ordered
phase stabilized in these films. 

A MnSi thin film sample was prepared by molecular beam epitaxy (MBE) on Si(111)
substrates as described in the supplemental information.\cite{SI}
\begin{figure}
\begin{center}
\epsfig{file=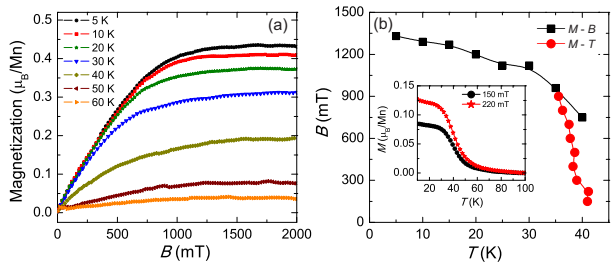,width=8cm}
\caption{Magnetometry measurements for the 40~nm-thick MnSi thin
  film. 
(a) Magnetization as a function of field at different
  temperatures.
(b) Phase boundaries 
identified by $B_{\mathrm{c}2}$ and $T_{\mathrm{c}}$.  Inset: Field-cooled magnetization curve in an applied of 150 and 220 mT.}
\label{SQUID}
\end{center}
\end{figure}
To characterize the film, a series of static magnetization
measurements was made by applying the magnetic field along the
MnSi[111] direction. The Curie temperature $T^{\mathrm{film}}_{\mathrm{c}}$, defined as the knee
point of the $M$-$T$ curve [examples shown inset Fig.~\ref{SQUID}(b)]
is found to be
$T_{\mathrm{c}}^{\mathrm{film}}=42.3(2)$~K,  which is consistent with the values 
previously reported for epitaxial MnSi
films.\cite{growth,karhu,wilson,karhu2,wilson2,li} The difference
between this value and that of
the bulk 
is largely attributable to tensile strain induced by the $-3\%$ lattice mismatch
between the MnSi film and the Si substrate, as discussed
previously.\cite{growth} 
The $M$-$B$ curves [Fig.\ref{SQUID}(a)] were measured at different temperatures after
field-cooling from 300~K in an applied field of 2~T. 
The Si substrate
provides a large diamagnetic background at low temperatures below
$T_{\mathrm{c}}$ which, along with contributions from unsaturated MnSi
moments at high field, \cite{unsat} results
 in a linear field dependence 
which we
subtract from the data. 
The saturation magnetization $M_{\mathrm{s}}$ at 5~K is found to be
0.41(3)$\mu_{\mathrm{B}}/\text{Mn}$, consistent with bulk
behavior\cite{unsat,bulk} but different from the behavior found in films
that are thinner than 10~nm.\cite{growth} 
The shape of the curves is consistent with previous
studies on epitaxial MnSi films with applied field directed out-of-plane.\cite{li,wilson2} 
This contrasts with 
field in-plane configuration (not shown), in which a sharp phase
transition can be identified at lower fields in the susceptibility
curves. (This effect appears to indicate that the system undergoes
different evolutions of the magnetic structure in the two distinct
geometries.) 
The kink in the hysteresis loops reported in Ref.~\onlinecite{li}
appears very weak, but can be seen in data measured between 600-800~mT
for 10~K, 20~K, and 30~K.

The transition from the conical phase to the
field-polarized phase at an applied field $B_{\mathrm{c}2}$ can be
extracted from the
$M$--$B$
curves at the point where $\text{d}M/\text{d}B=0$. 
The values of $B_{\mathrm{c}2}$ are found to be slightly higher than the intrinsic
values reported previously, and this may be caused by
demagnetization effects\cite{growth} due to the irregular sample shape
used in our magnetometry measurements. Figure~\ref{SQUID}(b) shows the
upper boundary of the phase diagram depicted by a series of
$T_{\mathrm{c}}$ values from $M$-$T$ curves, and $B_{\mathrm{c}2}$
values from $M$-$B$ curves, inside which the non-trivial spin textures
may appear. 

TF $\mu^{+}$SR measurements were made on the MnSi sample 
using the Low
Energy Muon (LEM) beamline at S$\mu$S.\cite{SI,prokscha}
Applied magnetic fields were directed perpendicular to the surface of the
sample (i.e. along [111]). 
Our use of TF $\mu^{+}$SR to probe the SL is analogous to its use in probing the  vortex lattice (VL) in
a type II superconductor, where the technique provides a powerful
means of measuring
the internal magnetic
field distribution caused by the presence of the magnetic field
texture.\cite{sonier} We have previously used this technique to
probe the SL region in bulk Cu$_{2}$OSeO$_{3}$.\cite{Lancaster-2015}
In a TF $\mu^{+}$SR experiment, spin polarized muons are implanted
in the bulk of a material in the presence of a
magnetic field $B_{\mathrm{app}}$, which is applied perpendicular
to the initial muon spin direction. Muons stop at
random positions on the length scale of the field texture where they
precess about the total local magnetic field $B$ at the muon
site, with frequency 
$\omega = \gamma_{\mu}B$, where $\gamma_{\mu} = 2 \pi \times
135.5$~MHz T$^{-1}$.
The observed property
of the experiment is the time evolution of the muon
spin polarization $P_{x}(t)$, which allows the determination of
the distribution $p(B)$
 of local magnetic fields across the
sample volume via $P_{x}(t) = \int_{0}^{\infty}\mathrm{d}B\, p(B) \cos
(\gamma_{\mu}Bt + \phi)$
where the phase $\phi$ results from the detector geometry. 

\begin{figure}
\begin{center}
\epsfig{file=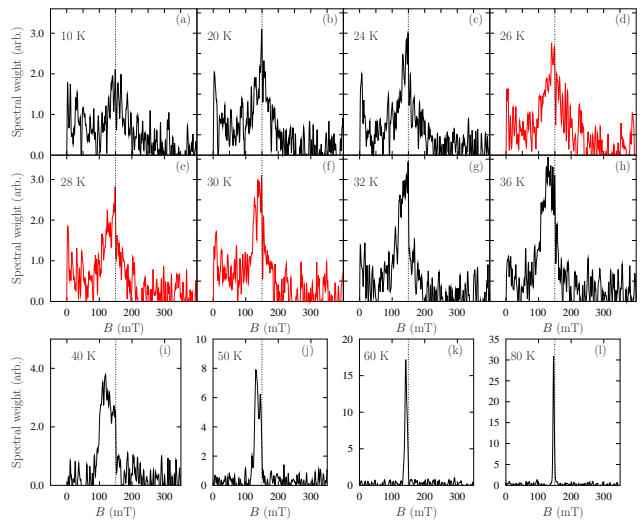,width=\columnwidth}
\caption{Fourier transform $\mu^{+}$SR  spectra measured for a MnSi
  film in an applied
  field of $B_{\mathrm{app}}=149$~mT for muons implanted into the center of the MnSi
  layer in the film. The skyrmion $A$-phase region in the bulk is
  found between $26 \lesssim T \lesssim 30$~K [spectra (d-f)]. Note the change of scale in
  panels (i-l). \label{data3}}
\end{center}
\end{figure}

Our TRIM.SP calculations\cite{SI} predict that an incident energy of $E=5$~keV
leads to fairly symmetrical implantation profile of muons, with a
maximum $\approx25$~nm below the surface of the 40~nm-thick MnSi layer with FWHM width $\approx 20$~nm, 
such that $>$95\% of muons are implanted in the MnSi film.  
TF $\mu^{+}$SR measurements were made in an applied magnetic field of $B_{\mathrm{app}}=149$~mT after
pre-cooling in this field. This field was chosen as it is known to promote the SL phase in the
bulk material. 
Example Fourier transform spectra, whose spectral weight is
proportional to $p(B)$, measured as a function of
temperature are shown in Fig.~\ref{data3}.
At high temperatures we observe the response of muons precessing in
the applied field $B_{\mathrm{app}}=149$~mT.
On cooling below 60~K the observed lineshape broadens
considerably, with its mean $B_{0}$
shifting to fields lower than $B_{\mathrm{app}}$. The
lineshape also becomes slightly skewed, with additional spectral weight shifting to
lower fields. 
The lineshape is seen to broaden further below the 30~K $A$-phase boundary observed
in the bulk, and again below 20~K. It is notable that the spectral lines
observed in our film sample seem somewhat less well resolved than
those seen in a
bulk sample.\cite{Amato-2014} This, along with the increased $T_{\mathrm{c}}$
observed in MnSi films, might be ascribable to strain effects in the
films.

\begin{figure}
\begin{center}
\epsfig{file=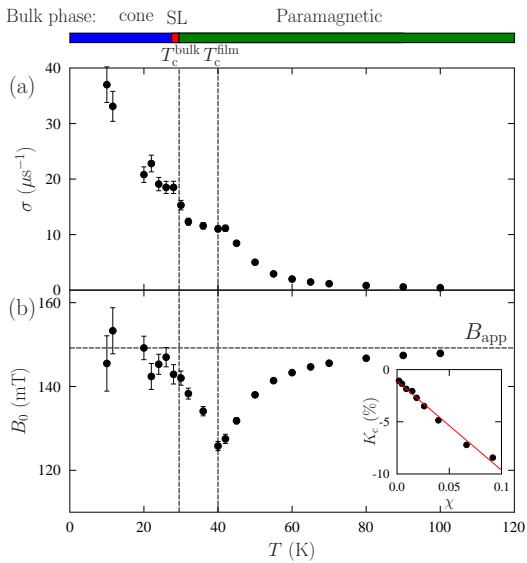,width=7cm}
\caption{Results of fitting the time domain spectra measured in
  applied field of $B_{\mathrm{app}}=149$~mT and implantation energy
  $E=5$~keV.
(a) Relaxation rate $\sigma$; (b) average magnetic
  field $B_{0}$. Dashed vertical lines show the positions of the critical
  temperatures found in films $T^{\mathrm{film}}_{\mathrm{c}}= 40$~K and
  the bulk $T^{\mathrm{bulk}}_{\mathrm{c}}=29.5$~K. The bar indicates the phases
  found in the bulk at 150~mT. The dashed horizontal line
  shows $B_{\mathrm{app}}$. Inset shows the muon Knight shift $K$ as a
  function of $\chi = \mu_{0}M/B$.\label{fig1}}
\end{center}
\end{figure}

The spectra were fitted in the time domain to a polarization function 
$P_{x}(t) = a{\rm  e}^{-\sigma^{2}t^{2}}\cos(\gamma_{\mu}B_{0}t+\phi),$
where $\phi$ are phases resulting form the detector geometry, 
and $a$ is a fixed amplitude. 
The evolution of (a) the relaxation rate $\sigma = \gamma_{\mu}\sqrt{\langle \left(  B- B_{0}
 \right)^{2} \rangle}/{\sqrt{2}}$, arising from those muons stopping
in MnSi, and (b) the average magnetic
field $B_{0}$ that they experience, are shown in Fig.~\ref{fig1}. 
The relaxation rate
$\sigma$ [Fig.~\ref{fig1}(a)] is seen to slowly increase with decreasing temperature below 100~K, with the rate
of increase becoming larger below 60~K. There is a cusp in the 
curve at 40~K, with the width remaining
approximately constant in the region between 40~K and 30~K.
At 30~K, the relaxation rate $\sigma$
jumps in magnitude, and  increases sharply once again below 20~K. Although a smooth increase
in $\sigma$ below $T^{\mathrm{film}}_{\mathrm{c}}$ might be predicted on the grounds
of an increase in the ordered moment of the system with decreasing
temperature in the ordered phase, the observed discontinuities are not be
expected in a typical ordered magnet. 
The average field $B_{0}$ [Fig.~\ref{fig1}(b)] is close to the applied
field in the high temperature regime around 100~K, but is seen to
decrease with decreasing temperature above
$T_{\mathrm{c}}^{\mathrm{film}}$, before reaching a sharp minimum
at 40~K. It then increases in the ordered regime until $B_{0}\approx
B_{\mathrm{app}}$ below $T\approx 30$~K.

A temperature scan carried out at a larger applied field of 220~mT (at the
same 5~keV
implantation energy) was found to show the same trends above 40~K.
The 
larger relaxation rates measured in these data
are more difficult to fit and consequently more scatter is seen at low
temperature, but the data are suggestive of similar features as those
seen at $149$~mT. Finally, a scan carried out
for $B_{\mathrm{app}}=149$~mT, but at a muon implantation energy
$E=1.3$~keV allows us to probe the response of muons implanted near to the surface of
the film. In this case similar features are again seen down to 40~K,
below
which the broad signal lineshapes become difficult to fit.\cite{SI} 

Each of the features seen in the TF $\mu^{+}$SR may be correlated with those observed
previously using other techniques in bulk and thin film samples of
MnSi.
We note that the general trend in $B_{0}$ above $T_{\mathrm{c}}^{\mathrm{film}}$ is
accounted for by  the
effect of the increase in magnitude of the hyperfine coupling with
decreasing temperature in the paramagnetic regime near the
transition, as has previously been observed in Knight shift
measurements.\cite{hayano} In fact, correcting the shift for the Lorentz
field and the demagnetizing field gives the Knight shift shown inset in
Fig.~\ref{fig1}, which when plotted against $\chi \approx \mu_{0}M/B$,
measured at $B=150$~mT, allows us to estimate a contact hyperfine
coupling of $A=-0.95(5)$~mol emu$^{-1}$, consistent with previous measurements.\cite{Amato-2014}

 In the bulk, the magnetic transition from the paramagnetic to the
skyrmion phase occurs in these applied fields at around\cite{bauer} $T^{\mathrm{bulk}}_{\mathrm{c}}=30$~K, where we see a jump in
the $\mu^{+}$SR relaxation rate $\sigma$ and change in behavior of the peak
field $B_{0}$. 
In addition, the film ordering transition $T_{\mathrm{c}}^{\mathrm{film}}$, identified from magnetometry and the onset of a sizeable response
from the THE \cite{li}, occurs at 40~K, where we see a knee in the evolution of
$\sigma$ and the maximum shift in $B_{0}$. 
Finally, THE measurements also show a transition or
crossover below the 15--20~K region, resulting in a small or negative
THE signal, which coincides with a further jump in
$\sigma$. We note that in the muon data the lineshape broadens
further still at low temperatures, possibly reflecting the increased
ordered moment size.

The original interpretation of the THE data\cite{li}  was based
on the assumption that a large response was the result of the
scattering of carriers from a topologically non-trivial spin
texture, identified with the SL phase. 
In view of the controversy surrounding the observation of skyrmions
using Lorentz TEM \cite{Monchesky-2014}, 
an alternative interpretation was suggested, \cite{meynell} which attributed the
topological contribution to the Hall effect to additional scattering of
charge carriers arising intrinsically in the magnetic cone phase. (It was
also noted that the lack of any clear transition being observed in the Hall effect
data prevented the signal being attributed to a separate magnetic
phase. \cite{meynell} However, it is also worth noting that 
although  the signal is dominated in many films by a feature in
the anomolous Hall signal that prevent the transition being observed, 
the transition can be seen in thinner films.)
Our results suggest that the magnetic properties of this system are unlikely to be
accounted for simply via effects arising from
a single magnetic
cone phase, but rather suggest
transitions in the nature of the field distribution occurring at
$T_{\mathrm{c}}^{\mathrm{film}}\approx 40$~K, $T^{\mathrm{bulk}}_{\mathrm{c}}\approx 30$~K and
possibly below 20~K.
Since the $T=30$ and 20~K features do not have a counterpart in the
magnetization measurements, they presumably do not involve a sizeable
rearrangement of the net component of the magnetization along the
field direction. 
However,  the changes in muon lineshape, reflected in the linewidth $\sigma$ and static
field shift in $B_{0}$, that we observe below $T_{\mathrm{c}}^{\mathrm{film}}$ are
suggestive of changes in the distribution of magnetic
fields at the muon sites.  This could suggest sizeable changes to the contact hyperfine
fields or to the nature of the dynamics. In the case of the latter,
for example, a freezing of relaxation channels in the fast-fluctuation
limit could lead to an increase in $\sigma$ on cooling. 
However, given the change in lineshape with temperature, the shift
in $B_{0}$ in the $30 \leq T \leq 40$~K region and the fact that a
transition around 30~K is known to occur in the bulk, 
we believe the most likely explanation
is that the features observed imply changes in the ordered
spin structure (which itself would likely also involve a change in
hyperfine fields and dynamics).

\begin{figure}
\begin{center}
\epsfig{file=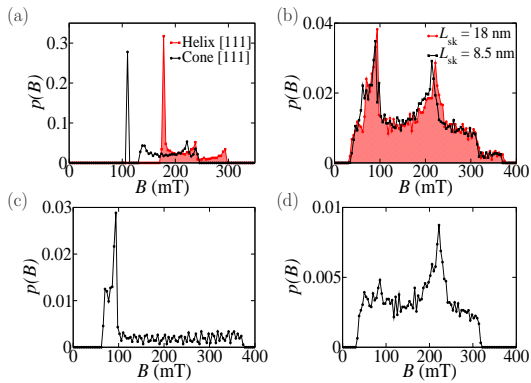,width=7.4cm}
\caption{ The $p(B)$ distribution in an applied field of
  $B_{\mathrm{a}}=150$~mT. 
(a) Helical and conical
  magnetic structures. 
(b) The $p(B)$ distribution for the SL magnetic structure with lattice constants
$L_{\mathrm{sk}}=8.5$~nm and 18~nm.
(c),(d) field distributions at stopping sites 0 and 1 respectively,
for the SL structure.
\label{theory}}
\end{center}
\end{figure}

To assess the evidence for the existence of skymions in this system, we have carried out simulations of the predicted muon lineshape in order to
compare the expected signal for a cone, helical and skyrmion lattice
phases. 
The local magnetic field distributions were generated following the procedure
outlined  in Ref.~\onlinecite{Lancaster-2015}.  
We  generate  helical and  conical  spin  textures
with propagation vectors oriented along the [111] direction. The SL is
generated with the skyrmion lattice plane perpendicular to [111]. 
Helical and SL  distributions are generated as described
previously\cite{Lancaster-2015}, while cone-phase distributions     are      generated
using \cite{wilson2}
$\boldsymbol{m} (\boldsymbol{r})/\boldsymbol{m}_{0}=\hat{\boldsymbol{x}}\sin\theta\cos(\boldsymbol{q}\cdot\boldsymbol{r}) +
                 \hat{\boldsymbol{y}}\sin\theta\sin(\boldsymbol{q}\cdot\boldsymbol{r}) +
                 \hat{\boldsymbol{z}}\cos\theta,$
where $\cos\theta=|B|/B_{\mathrm{c}2}$ and  $\boldsymbol{B}\parallel\hat{\boldsymbol{z}}\parallel [111]$.  
 The  magnetic moment is taken  to be
0.4$\mu_B$   \cite{Motoya-1978},    $B_{\mathrm{c}2}=0.5$~T
(corresponding to  $T_{\mathrm{c}}^{\mathrm{film}}\approx 40$~ K  \cite{wilson2})
and the length scale of the  helical
\cite{Amato-2014} and conical structures is taken to be 18~nm. 
For the skymion lattice  we consider lattice constants suggested in
previous studies of both
$L_{\mathrm{sk}}= 18$~nm\cite{tonomura} and 8.5~nm\cite{li}, although
there is little difference between the two.
We evaluate dipole fields from the moment distributions along
with the contribution from the contact hyperfine field, calculated using \cite{Amato-2014}
$
B_{c} = \frac{4 V_{\mathrm{mol}}A_{\mathrm{HF}}}{NV_{\mathrm{cell}}}\sum_{j=1}^{N}\boldsymbol{m}(\boldsymbol{r}_{j}),
$
where $V_{\mathrm{mol}}$ is the molar volume of Mn ions,
$V_{\mathrm{cell}}$ is the unit cell volume,
$A_{\mathrm{HF}}=-0.9276$~mol/emu is the hyperfine coupling and $N$ is the number of
Mn ions within one lattice constant of the muon site. 
The field distribution $p(B)$ sampled by the muon ensemble has been
shown to arise from
muons stopped in the $4a$ Wyckoff position. These
give rise to two magnetically distinct classes of site, conventionally
labelled sites 0 and 1.\cite{Amato-2014}

The predicted field distributions $p(|B|)$ in an applied field of
$B_{\mathrm{app}} =150$~mT, shown in Fig.~\ref{theory}(a,b), are
seen to be quite distinct in each phase.
The cone magnetic structure [Fig.~\ref{theory}(a)] involves a distribution of
fields (resulting from muons at site 1) and a sharp peak on the
low field side of the applied field (resulting from site 0). 
The signal from the skyrmion magnetic structure  [Fig.~\ref{theory}(b)] shows significant broadening over that resulting from the
cone and helical magnetic structures. The shape remains asymmetric, but is less
skewed than the distribution expected from the conical structure. 
In addition, the contribution from 
site 0 shows an asymmetric distribution, with a sharp peak on the
low field side of the applied field. 
The features in the simulations are not apparent in the measured data,
which also includes, for example, a background contribution from
muons stopping in the sample holder. 
 We note, however, that 
spectra measured in the region where the SL phase is
observed in bulk MnSi ($26 \lesssim T
\lesssim 30$~K) 
do show a broadened field distribution compared to those measured in
the $30 \lesssim T \lesssim 40$~K region, with some spectral weight shifted to
fields below the diamagnetic peak. 
It is also noticable that $p(B)$ for the cone structure involves a sizeable shift of the peak in spectral
weight to low fields and that, in the data, the value of the field $B_{0}$
recovers towards the applied field below 40~K with the shift no longer
apparent below $T=30$~K.
Despite these observations, it is not
possible from the comparison of these simulations to the measured
lineshapes of Fig.~\ref{data3}, to make firm conclusions regarding the nature of the spin
structure in the films, nor to unambiguously conclude whether skyrmions
are present. 
Whatever the case, 
it is unlikely that the changes that we see in the $T=20$--$40$~K region can be accounted for
simply by a single magnetic cone phase. 
It is also unlikely that our measurements could be explained via
a single
magnetic phase with the formation of chiral domains of
the form discussed, e.g., in Ref.~\onlinecite{marrows}. In that case, we
would expect the muon response in the two domains to be 
different (based on the behaviour of the bulk material \cite{Amato-2014}), resulting in a broadening compared to the
single domain case. However this would not, in itself, explain the succession of 
changes in broadening we observe on cooling. 

In conclusion, magnetization and 
TF $\mu^{+}$SR measurements in thin film MnSi 
identify an ordering temperature at
$T_{\mathrm{c}}^{\mathrm{film}}\approx 40$~K. 
The TF data also
reveal significant changes in the
static field distribution that coincide with the topological
contribution to the Hall effect identified previously, and also with
the the magnetic phase boundaries observed in both the bulk and thin
films. We therefore suggest that there may be phase boundaries or
crossovers in behaviour in this system
occurring around $T\approx 20$~K, 30~K and 40~K.
Although our data do not reveal a signature of the
SL phase, it is unlikely that a single cone phase 
could account for our results. 

This work was carried out at S$\mu$S, Paul Scherrer Institut,
Switzerland and 
we are grateful
for the provision of beam time. 
We thank the EPSRC (UK) and the John Templeton Foundation 
for financial support and to A. Amato, Ch. Pfleiderer and R.C. Williams for useful discussion.
SLZ and TH gratefully acknowledge support by the Semiconductor Research Corporation (SRC).
This work made use of the facilities of N8 HPC provided and funded by
the N8 consortium and EPSRC (Grant No.\ EP/K000225/1). The Centre is
co-ordinated by the Universities of Leeds and Manchester.

\end{document}